\documentclass{aa}
\usepackage{amsmath}
\usepackage[svgnames]{xcolor}
\definecolor{blue}{RGB}{0,0,255}
\usepackage{caption}
\usepackage[font={color=blue},figurename=Fig.,labelfont={it}]{caption}
\usepackage{bm}
\usepackage{subcaption}
\usepackage[varg]{txfonts}
\usepackage{graphicx}
\usepackage{natbib}
\usepackage{pdfpages}
\usepackage{wasysym}
\usepackage{rotating}
\bibpunct{(}{)}{;}{a}{}{,} 
\usepackage{hyperref}
\usepackage{soulutf8}
\usepackage{indentfirst}

\hypersetup{
    colorlinks=true,                          
    linkcolor=blue, 
    citecolor=blue, 
    urlcolor=blue,
    linktoc=all
}

\begin{document}
\title{On the annual and semi-annual components of variations in extent of Arctic and Antarctic sea-ice}
\author{F. Lopes \inst{1}
\and V. Courtillot \inst{1}
\and D. Gibert\inst{2}
\and J-L. Le Mouël\inst{1}
}
\institute{{ Universit\'e de Paris, Institut de Physique du globe de Paris, CNRS UMR 7154, F-75005 Paris, France}
\and { LGL-TPE - Laboratoire de G\'eologie de Lyon - Terre, Planètes, Environnement, Lyon, France}
}
\date{}

\abstract {The time series (1978-2022) of northern hemisphere (\textbf{NHSI}) and southern hemisphere (\textbf{SHSI}) sea-ice extent are submitted to singular spectral analysis (\textbf{SSA}). The spectral components are analyzed with Laplace’s formulation of the Liouville-Euler system of differential equations.  Laplace assumes that all masses on and in Earth are set into motion by astronomical forces. As already shown in a previous work (see {\textcolor{blue}{Le Mouël et al., (2021b)}}, figure 03), the trends observed in the time series are quasi linear, decreasing for \textbf{NHSI} (by 58.300 km$^2$/yr) and increasing for \textbf{SHSI} (by 15.400 km$^2$/yr). The amplitude of annual variations in the Antarctic is double that in the Arctic, they are in phase opposition and modulated, with a flat trend for \textbf{SHSI}, a decreasing trend for \textbf{NHSI} and a sub-decadal modulation of the envelopes. The semi-annual components are in quadrature. The first three oscillatory components of both \textbf{NHSI} and \textbf{SHSI} at 1, 1/2 and 1/3 yr account for more than 95\% of the signal variance.  The trends are respectively 21 (Antarctic) and 4 times (Arctic) less than the amplitudes of the annual \textbf{SSA} components. Following Laplace’s views, we complement previous analyses of variations in pole position (\textbf{PM} for polar motion, with coordinates $m_1$, $m_2$) and length of day (\textbf{lod}). Whereas \textbf{SSA} of \textbf{lod} is dominated by the same first three components as sea-ice, \textbf{SSA} of  \textbf{PM} contains only the 1 yr forced annual oscillation and the Chandler $\sim$1.2 yr components. The 1 yr component of \textbf{NHSI} is in phase with that of lod and in phase opposition with m1. The reverse holds for the 1 yr component of \textbf{SHSI}. We note that the semi-annual component appears in \textbf{lod} not in $m_1$. The annual and semi-annual components of \textbf{NHSI} and \textbf{SHSI} are much larger than the trends observed since 1978, that leads us to test whether a first order geophysical or astronomical forcing should not be preferred to the mechanisms generally suggested as a forcing factor of the trends. The lack of modulation of the largest (\textbf{SHSI}) forced component suggests an alternate mechanism. In Laplace’s paradigm, the torques exerted by the Moon, Sun and planets play the leading role as the source of forcing (modulation) of many geophysical phenomena. These forces (and torques) lead to changes in the inclination of the Earth’s rotation axis, transferring stresses to the Earth’s solid and fluid envelopes, setting Earth masses in motion and resulting in thermal dissipation: more than variations in eccentricity, it is variations in inclination of the rotation axis that lead to the large annual components of melting and re-freezing of sea-ice.}

\titlerunning{Variations in extent of Arctic and Antarctic sea-ice}
\maketitle

\section{Introduction} 
	In a series of previous papers, we have applied singular spectrum analysis (\textbf{SSA}) to long time series of geophysical observables. We have analyzed series for which several decades to centuries of data were available, allowing one to explore their periodic or quasi-periodic components in the day to century time scales. We have successively analyzed polar motion and length of day \citep{LeMouel2019a,Lopes2021},  oceanic climate indices \textbf{MJO}, \textbf{PDO}, \textbf{ENSO}, \textit{etc} \ldots \citep{Courtillot2013,LeMouel2019b}, global surface temperatures \citep{LeMouel2020a}, and in order to contribute to the understanding of solar-terrestrial relationships sunspot numbers ISSN \citep{Courtillot2021}. The determination of a comprehensive range of pseudo-periodic components of \textbf{ISSN} has even allowed us to propose a prediction of the ongoing solar Cycle 25. In the present paper, we extend the same kind of analysis to an important component of the climate system, namely the variations in surface extent of sea-ice in the Arctic and Antarctic. \\	
	
In a previous paper on which the present study is also based, \cite{Lopes2021} recalled the importance of the treatise of celestial mechanics of \cite{Laplace1799}, and in particular the system of equations, later named after Euler and Liouville, that express the conservation of kinetic momentum and govern the declination and inclination of the rotation axis of any planet in revolution about the Sun (see Appendices {\textcolor{blue}{A}} and {\textcolor{blue}{B}}). Laplace’s paradigm can be summarized by two statements: (1) all masses on the surface or internal to a planet are set in motion by astronomical forces (particularly from the Sun, Moon and the Jovian planets), and (2) it is sufficient to know the polar motion or its time derivative (the \textbf{lod}) to have access to the periods of all moving masses on the surface or inside Earth. This is why the revolution periods of planets, or their commensurable combinations, and luni-solar tides have been found in many observed time series of geophysical phenomena, ranging from hours to millions of years \citep{Milankovic1920, MorthSchlamminger1979, Fairbridge1984, Morner1984, Laskar2004, Scafetta2010, Barnhart2011, Manzi2012, Scafetta2012, Morner2013, Scafetta2013, Lopes2017, Boulila2018, LeMouel2019a, LeMouel2019b, Dumont2020,LeMouel2020a, Lopes2021, Petrosino2022}. \\

Most recently, \cite{Courtillot2021} have shown that the sunspot number ISSN, a proxy of solar activity, could be decomposed in a series of oscillations, all with periods similar to or commensurable with those of Jovian planets \citep{MorthSchlamminger1979}. It has been possible to reconstruct more than 90\% of the variance of the original time series \citep{Courtillot2021}. Similar results have been obtained for polar motion, 90\% of the variance being carried by the ephemerids of the Jovian planets \cite{Lopes2021}. Pioneering work by \cite{Milankovic1920} and more recently by \cite{Laskar2004} explains how these planets influence over very long times the obliquity, precession and eccentricity of our planet’s orbit and rotation axis, and therefore among other consequences climate. \\

Planets exert gravitational forces (attraction) on all celestial bodies and because of its mass the Sun is the leading attractor (it exerts a force F of 3.57*10$^{22}$ kg.m.s$^{-2}$ on Earth). But because it is close to the center of gravity of the solar system its contribution to the budget of angular momenta (F*Sun-Earth distance*Earth’s revolution period = 1.68*10$^{41}$ kg.m$^{2}$.s$^{-1}$) is one to two orders of magnitude smaller than that of Jovian planets. In a heliocentric reference frame, Jupiter’s angular momentum  is 1.93*10$^{43}$ kg.m$^{2}$.s$^{-1}$, Saturn’s is 7.82*10$^{42}$ kg.m$^{2}$.s$^{-1}$, Neptune’s  2.50*10$^{42}$ kg.m$^{2}$.s$^{-1}$, and Uranus’s 1.69*10$^{42}$ kg.m$^{2}$.s$^{-1}$ . \\

We present the data for the North and South hemisphere sea-ice extents (\textbf{NHSI} and \textbf{SHSI} respectively) in section 2 and their spectral components using both Fourier analysis and \textbf{SSA} in section 3. We provide complementary results on polar motion and \textbf{lod} in section 4. We compare the annual components of polar motion and \textbf{lod} with those of sea-ice extent in section 5. A hypothesis is discussed in section 6 and we close with a conclusion in section 7.\\

\section{Available data for sea-ice extent in both hemispheres}
    Very precise daily series of sea-ice extent in the Arctic (\textbf{\textbf{NHSI}}) and Antarctic (\textbf{\textbf{SHSI}}) are available, starting in 1978 \footnote{\scriptsize ftp://sidads.colorado.edu/pub/DATASETS/NOAA/G02135/south/daily/data}. The data set \citep{Cavalieri2012, Fetterer2017} from which the \textbf{\textbf{NHSI}} and \textbf{\textbf{SHSI}} sea-ice areas are calculated consists of sea ice concentration maps derived from the radiances obtained from microwave radiometers on a suite of satellites. This started with the Nimbus 7 Scanning Multichannel Microwave Radiometer (\textbf{SMMR}), which operated from 1978 through 1987, then the Defense Meteorological Satellite Program (\textbf{DMSP}) series of F8, F11, F13 and F15 Special Sensor Microwave Imagers (\textbf{SSMI}), and the F17 Special Sensor Microwave Imager Sounder (\textbf{SSMIS}). The F8 operated from 1987 through 1991, the F11 from 1991 through 1995, the F13 from 1995 through 2007, and the F17 from 2008 through 2010 (\cite{Cavalieri1997,Cavalieri1999,Cavalieri2011}, \cite{Cavalieri2012}). After 2010 and up to 2018, we refer to \cite{Parkinson2019}, quoting  \cite{Cavalieri1999, Cavalieri2011}. \\
    
    The data set is generated using the Advanced Microwave Scanning Radiometer - Earth Observing System (\textbf{AMSR-E}) Bootstrap Algorithm, with daily varying tie-points. Daily (every other day prior to July 1987) and monthly data are available for both the north and south polar regions. Data are gridded on the SSM/I polar stereographic grid (25 x 25 km) and provided in two-byte integer format. The sampling from 10/26/1978 to 12/02/1987 is not as continuous as afterwards; there is only one datum every other day from 10/26/1978 to 08/20/1987, one per day from 08/20/1987 to 12/02/1987, then a gap from 12/02/1987 and 01/13/1988, back to one point per day from 01/13/1988 to 02/19/2022, a two day gap from 02/19/2022 to 02/22/2022, and finally full sampling (one per day) until 03/21/2022. We have worked with the data set as described above,  starting in 10/26/1978 and ending in 03/21/2022 (\href{https://nsidc.org/data/G02135/versions/3}{https://nsidc.org/data/G02135/versions/3}). The data are shown in Figure (\ref{Fig:01a})  (\textbf{NHSI} in blue, \textbf{SHSI} in red).\\

    A number of features are immediately apparent from this figure: annual variations are dominant, their amplitude in the Antarctic is double that in the Arctic, they are in phase opposition and modulated, with a flat trend for \textbf{SHSI}, a decreasing trend for \textbf{NHSI} and a sub-decadal modulation of the envelopes of both. Figure (\ref{Fig:01b}) shows the Fourier spectra of the two series with successive enlargements: a fundamental annual period is accompanied by 7 harmonics from 1/2 to 1/8 yr. \\
 \begin{figure}[h!]
     \centering
	\begin{subfigure}[b]{\columnwidth}
        \centerline{\includegraphics[width=\columnwidth]{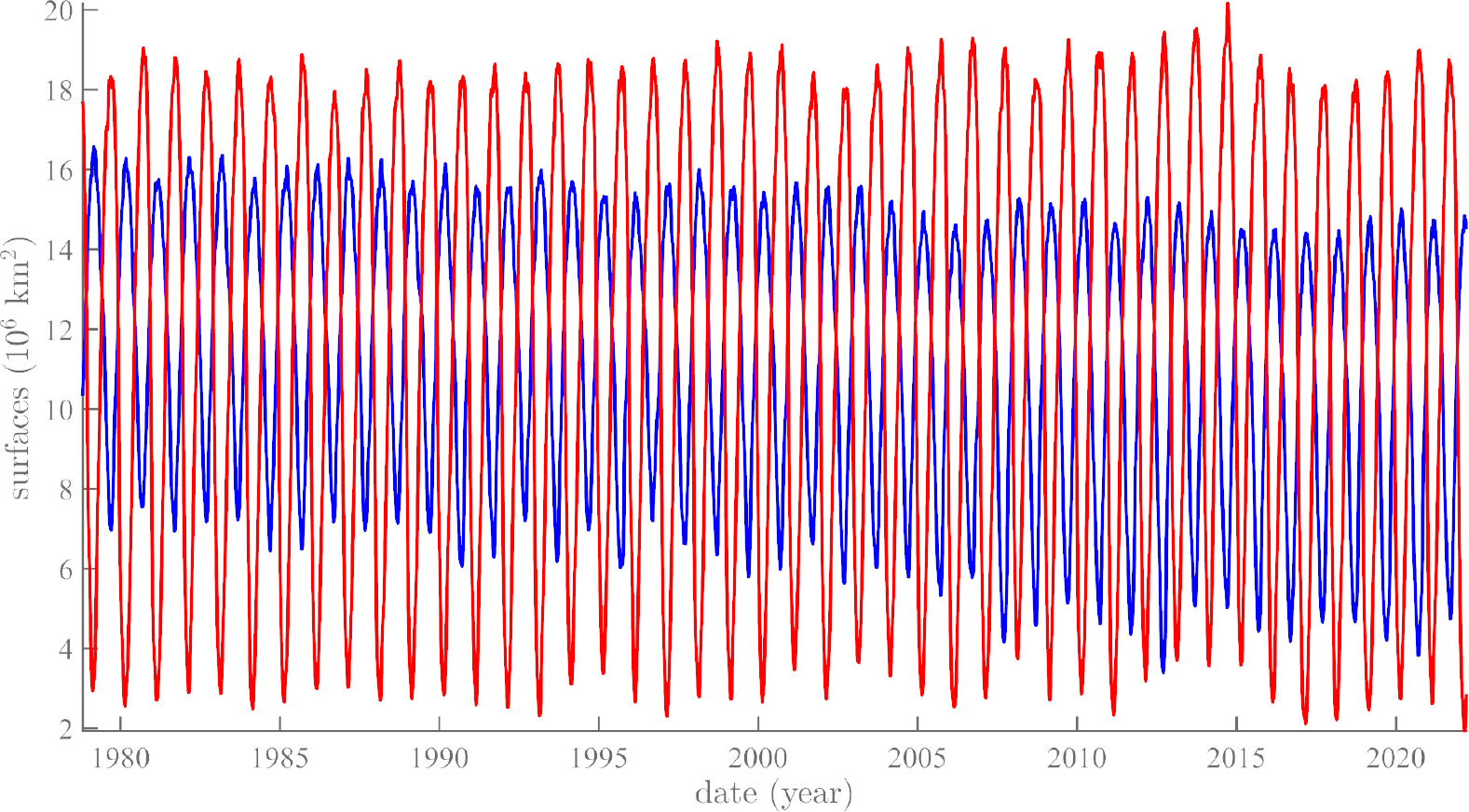}} 	
        \caption{Raw southern hemisphere (\textbf{SHSI}, red) and northern hemisphere (\textbf{NHSI}, blue) sea-ice data (https://nsidc.org/data/G02135/versions/3).}
        \label{Fig:01a}
	\end{subfigure}
\end{figure}
\begin{figure}[htb]\ContinuedFloat
    \centering
	\begin{subfigure}[b]{\columnwidth}    
	    \centerline{\includegraphics[width=\columnwidth]{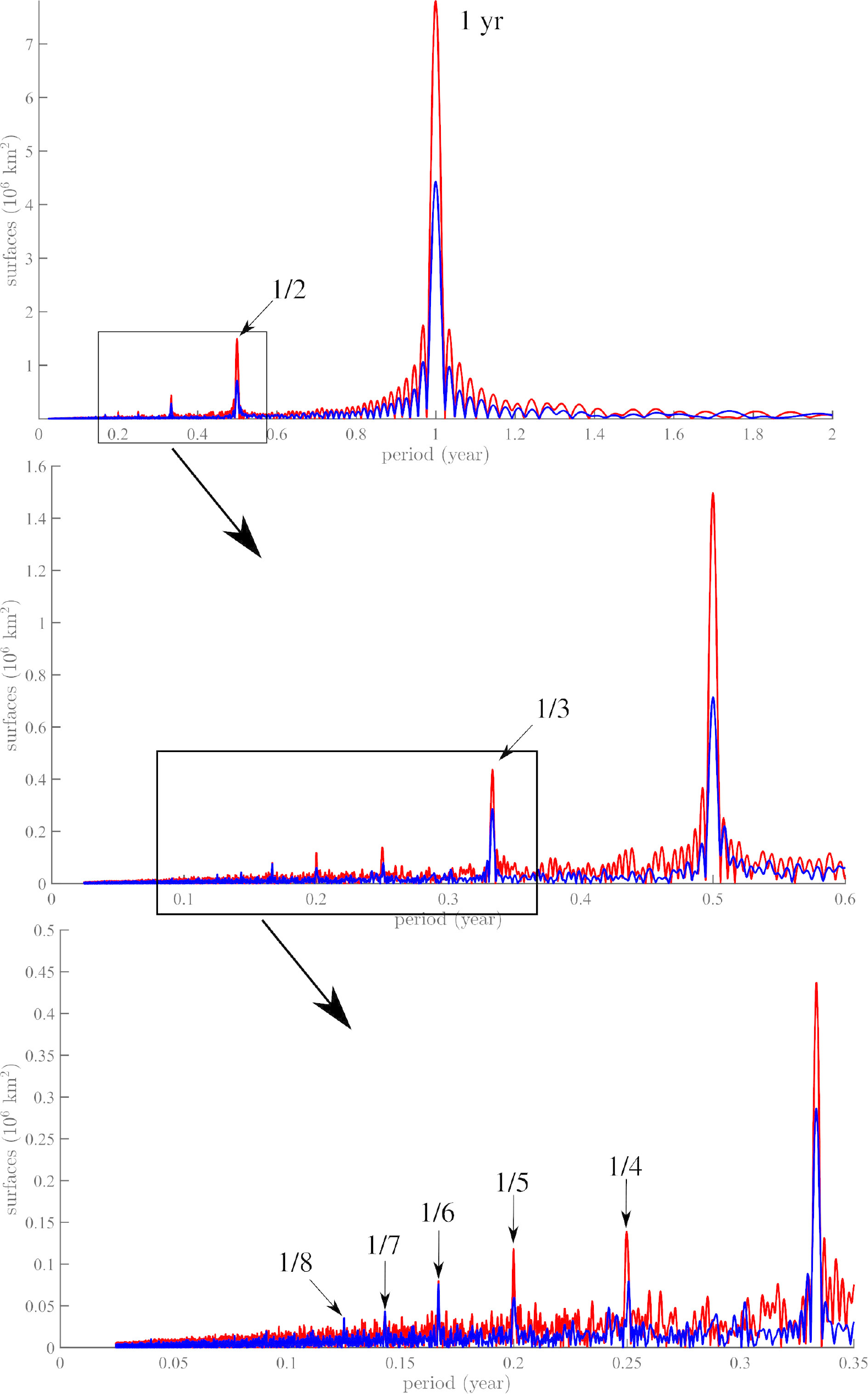}} 	
        \caption{Fourier spectrum of the \textbf{SHSI} (red) and \textbf{NHSI} (blue) series of Figure (\ref{Fig:01a}).}
        \label{Fig:01b}
    \end{subfigure}
	\caption{Temporal evolution of  \textbf{SHSI}(red) and \textbf{NHSI} (blue)}
	\label{Fig:01}
\end{figure}
\ \\
\newpage

\section{SSA analysis of the sea-ice extent data \textbf{NHSI} and \textbf{SHSI}}
    We next apply \textbf{SSA} (Singular spectral analysis; \cite{Golyandina2018}) as in our previous papers (\textit{e.g.} \cite{Lopes2021})  in order to determine the waveforms (amplitude and phase) of the periodic or quasi-periodic components of the sea-ice extent series. As seen in Figure \ref{Fig:01}, they are dominated by the modulation of a quasi-monochromatic cycle. \\

    For \textbf{NHSI}, the first three components allow one to recover 98.7\% of the total signal variance. They occur at periods 1.000 $\pm$ 0.014 yr (95.6\% of the variance), 0.499  $\pm$ 0.003 yr (2.69\%), and 0.333 $\pm$ 0.001 yr (0.43\%). For \textbf{SHSI} the first three components allow one to recover 97.9\% of the total signal variance. They occur at periods 0.999 $\pm$ 0.014 yr (94.4\%), 0.500  $\pm$ 0.004 yr (3.25\%), and 0.333 $\pm$ 0.001 yr (0.25\%). Remaining components have a negligible contribution. In the following, we focus on the annual and semi-annual components that are displayed in Figure (\ref{Fig:02a}) and(\ref{Fig:02b})respectively. \\
           
    The annual Antarctic sea-ice component is only slightly modulated on a multi-decadal time scale, whereas the annual Arctic component is modulated, with a significant growth. The two are in phase opposition (180 $\pm$ 1 day) as expected.  For the semi-annual components, the phase difference is 34 $\pm$ 2 days, which is close to quadrature (90/2 days). \\
    
    The trends are quasi linear (not shown in the figures), decreasing for \textbf{NHSI} (by 58.300 km$^2$/yr) and increasing for \textbf{SHSI} (by 15.400 km$^2$/yr). The total surface change since 1978 is a gain of 0.57*10$^6$ km$^2$ for the Antarctic and a loss of 2.15*10$^6$ km$^2$ for the Arctic. These values are in good agreement with \cite{Parkinson2019}. Note that they are respectively 21 (Antarctic) and 4 times (Arctic) less than the amplitudes of the annual \textbf{SSA} components.    
\begin{figure}[htb]	
    \centering
    \begin{subfigure}[b]{\columnwidth}
		\centerline{\includegraphics[width=\columnwidth]{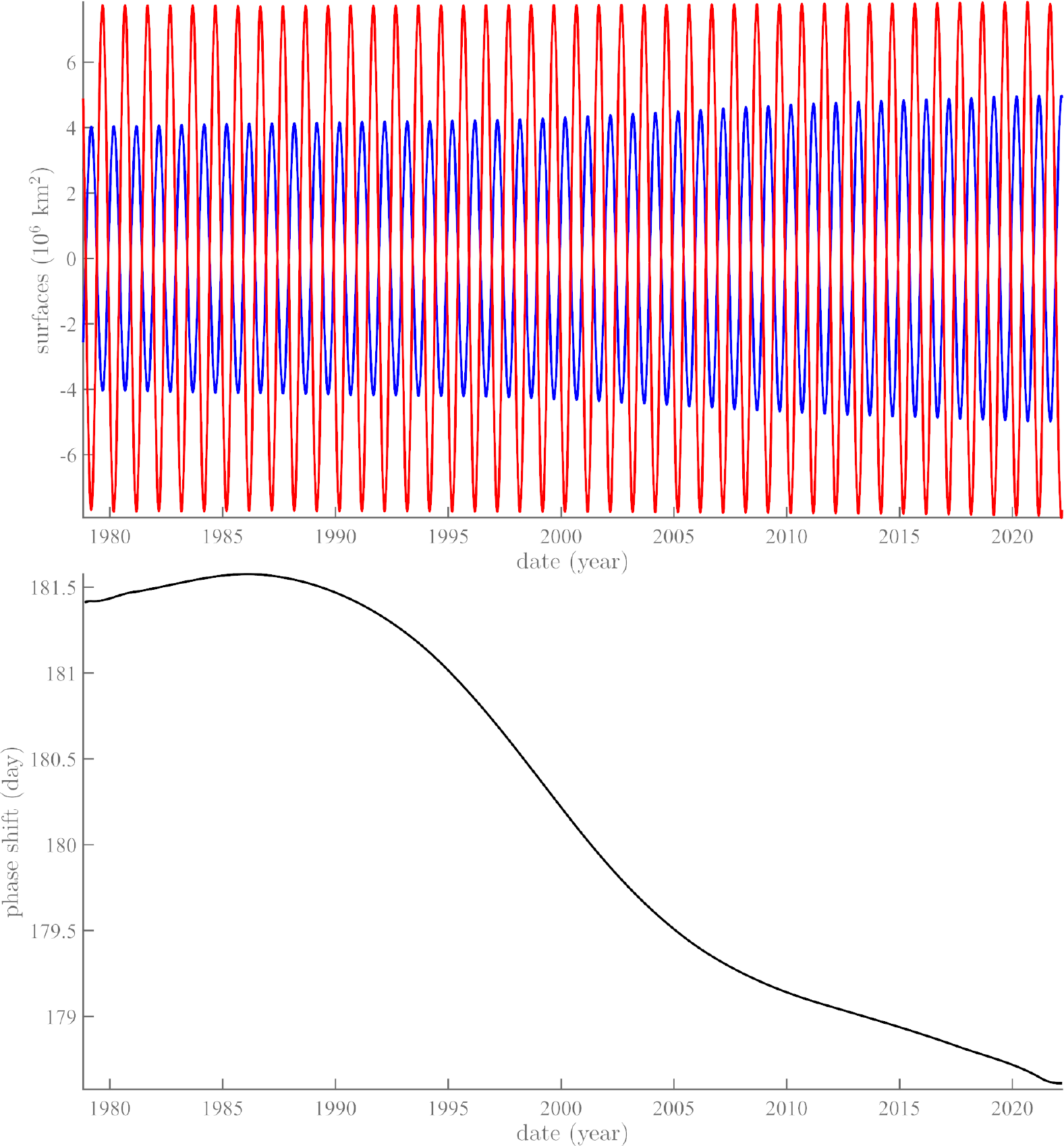}} 	
         \caption{SSA annual component of \textbf{SHSI} (red) and \textbf{NHSI} (blue) from 1978 to 2022 (top) and their phase difference (bottom; in days).}
		\label{Fig:02a}
	\end{subfigure}
\end{figure}	
\begin{figure}[htb]\ContinuedFloat
    \centering
	\begin{subfigure}[b]{\columnwidth}
		\centerline{\includegraphics[width=\columnwidth]{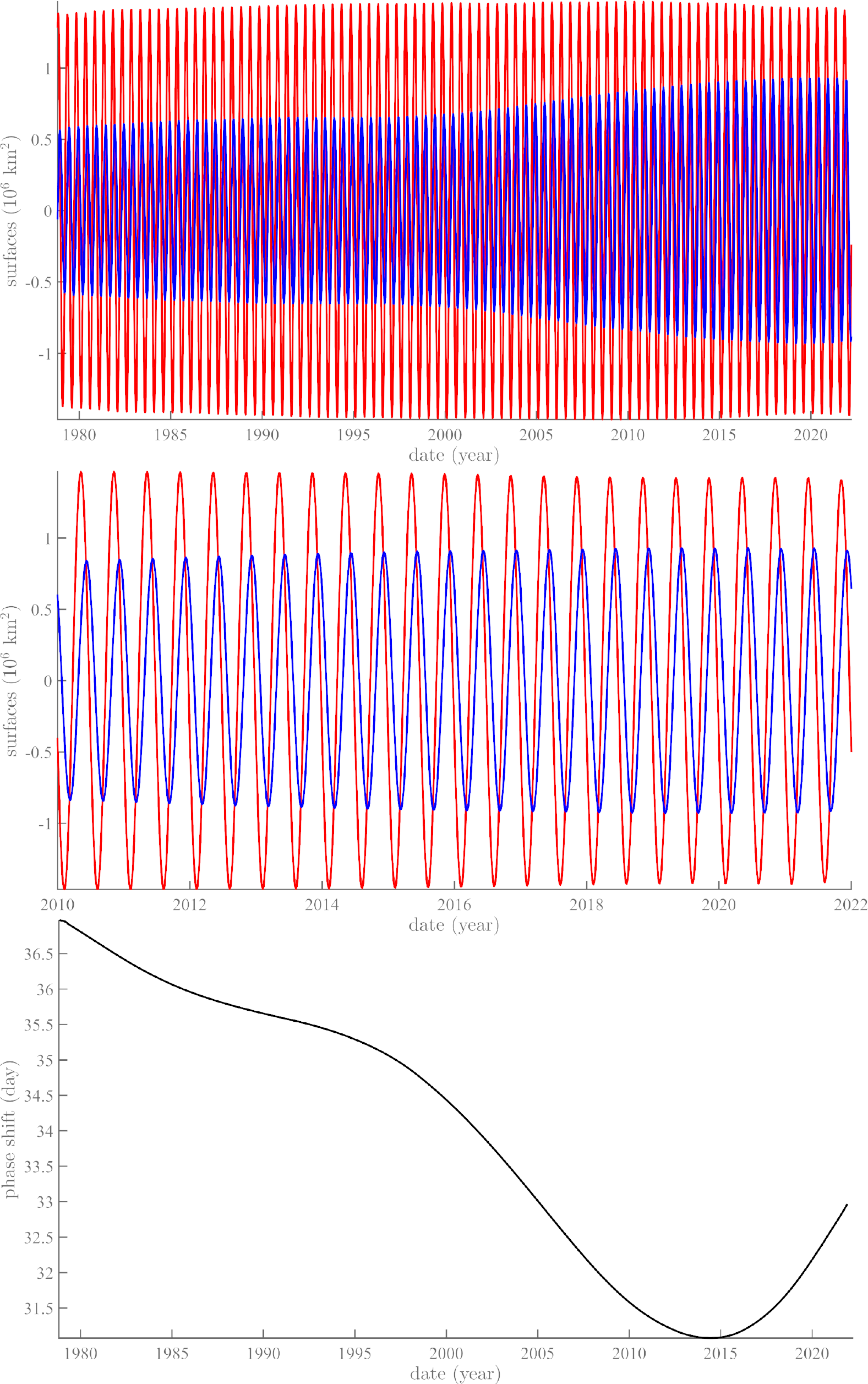}} 	
          \caption{SSA semi-annual component of \textbf{SHSI} (red) and \textbf{NHSI} (blue) from 1978 to 2022 (top), an enlargement from 2010 to 2022 (middle) and their phase difference (bottom, in days).}
		\label{Fig:02b}
	\end{subfigure}
	\caption{Annual and semi-annual components extracted from \textbf{SHSI}(red) and \textbf{NHSI} (blue)}
	\label{Fig:02}
\end{figure}    
 \newpage
\section{Complementary results for polar motion and length of day}
\begin{figure}[h]
        \centerline{\includegraphics[width=\columnwidth]{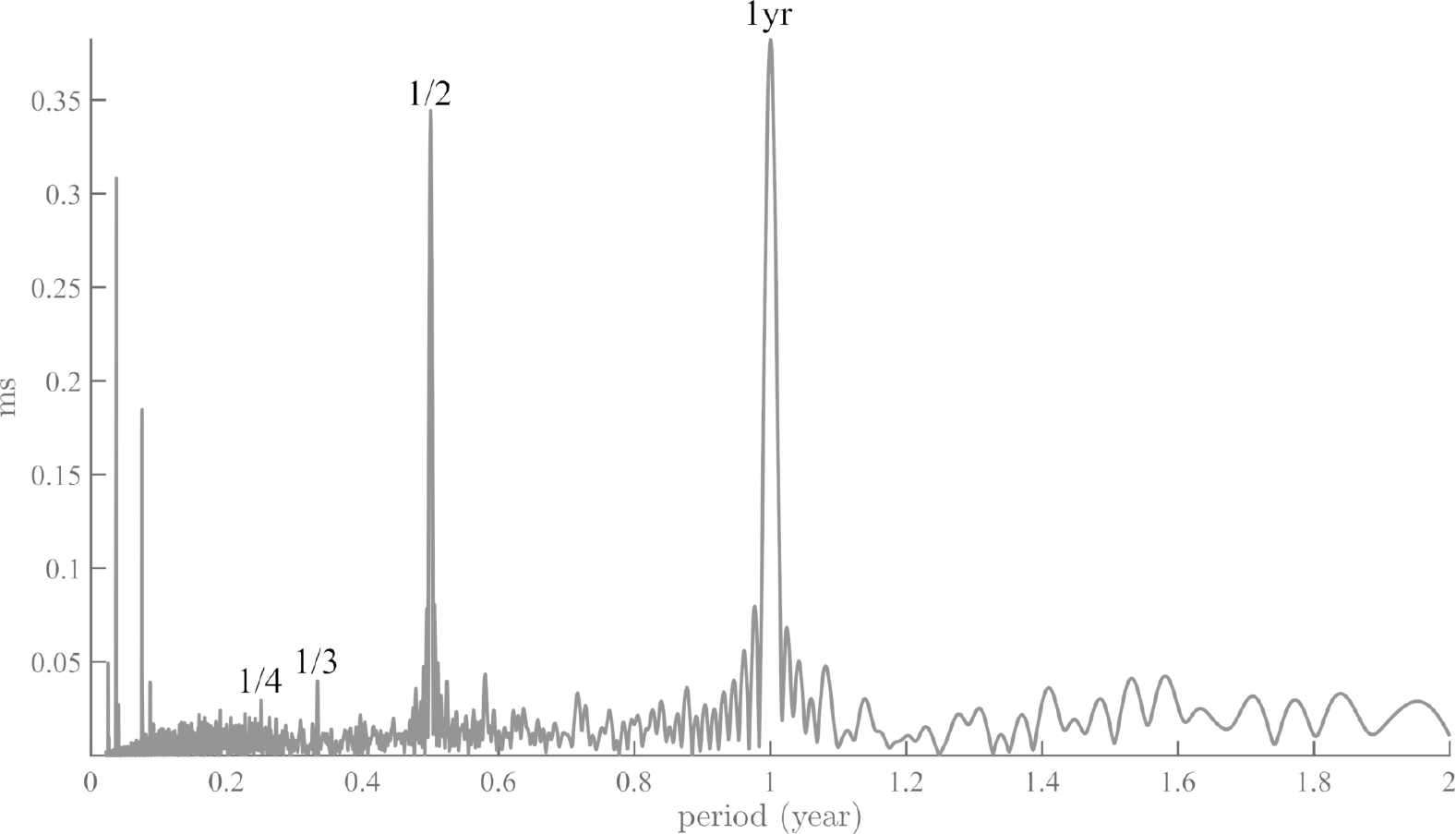}} 	
        \caption{Fourier spectrum of the \textbf{lod} time series (1962-2021).}
        \label{Fig:03}
\end{figure}

The orientation of the Earth’s rotation axis is on many time scales a key variable that modulates the insolation of any location on the Earth’s surface. It is therefore interesting to compare the spectral components of sea-ice extent with the variations in polar motion (\textbf{PM}) and length of day (\textbf{lod}). These data are available on the site of the International Celestial Reference System (\textbf{IERS})\footnote{\scriptsize{https://www.iers.org/IERS/EN/DataProducts/EarthOrientationData/eop.html}}, where we have selected file EOP14C04. Several authors have studied the spectral content of \textbf{lod}. In \citep{LeMouel2019a} we explored with \textbf{SSA} the range from 9.13 days to 18.6 years. Longer periods at 18.6 yr, 11 yr, and some harmonics, and shorter periods at 27 days and harmonics are linked to the Sun and Moon, and are discussed elsewhere \citep{LeMouel2019a}. For this paper, we focus on the range below 2 years (Figure \ref{Fig:03}; Fourier spectrum). Distinct peaks are seen at 1 and 1/2 yr, but also at 1/3 and 1/4 yr. The leading two terms amount to 13.25\% and 11.06\% respectively for a total of 24.3\% of the total \textbf{lod} variance. \\    

\begin{figure}[h]
        \centerline{\includegraphics[width=\columnwidth]{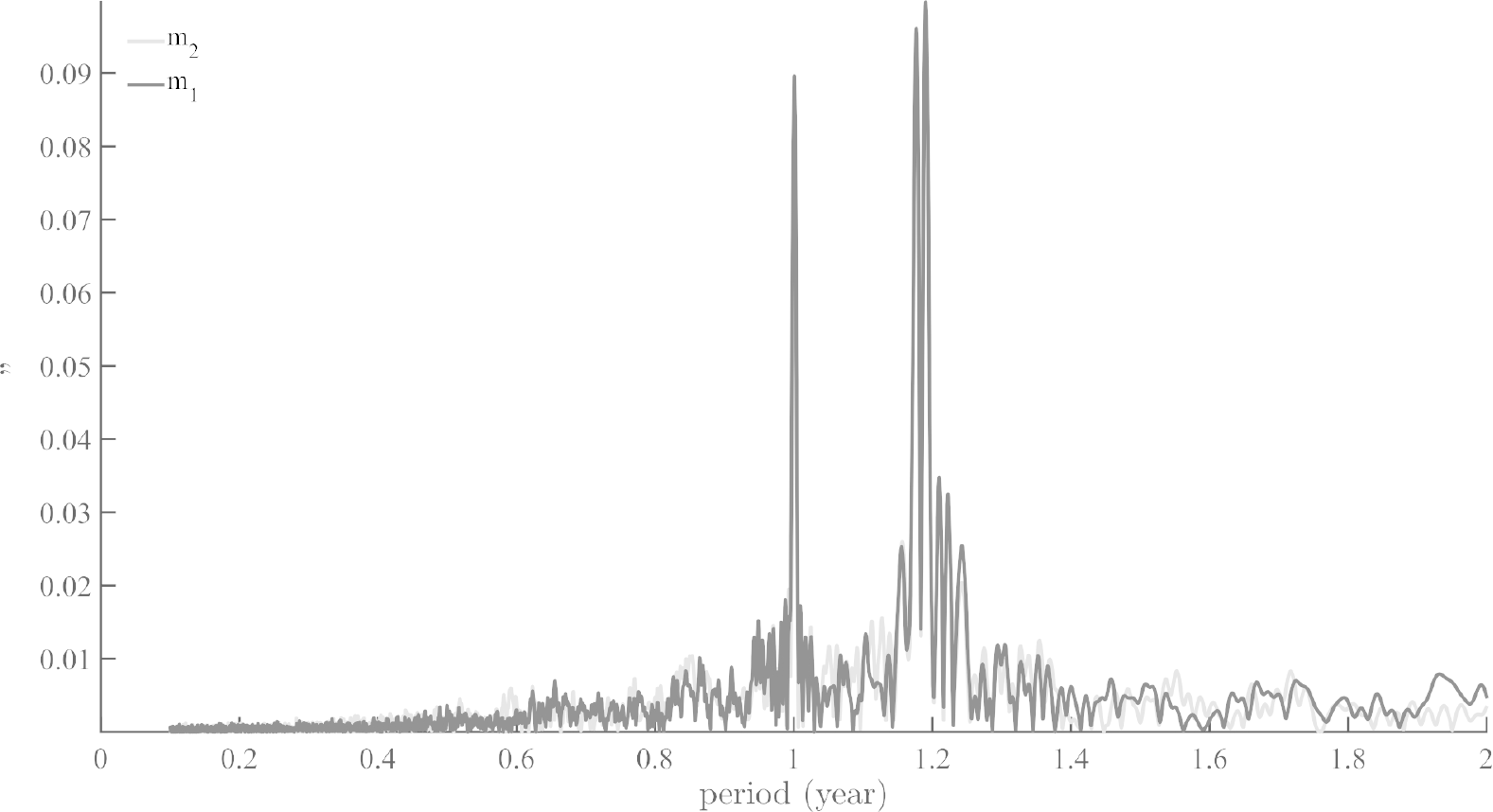}} 	
        \caption{Fourier spectrum of the coordinates of the rotation pole ($m_1$, $m_2$) time series (1978-2022)}
        \label{Fig:04}
\end{figure}

Let us turn to the time series (1978-2022) of coordinates of the pole of rotation ($m_1$, $m_2$ ; see \cite{Lopes2022a}). The Fourier analysis of the modulus \textbf{m} (hereafter \textbf{PM}, for polar motion) is shown in Figure \ref{Fig:04}. The spectrum features only two sharp peaks at 1 yr and 1.2 yr. Components with decadal to ~80 yr (Gleissberg cycle) periods are present but out of the picture, and so is the trend known as the Markowitz drift (\cite{Markowitz1968}, \cite{LeMouel2017, LeMouel2021}). The 1/3 yr and higher harmonics seen in \textbf{NHSI}, \textbf{SHSI}, and \textbf{lod} are not found in \textbf{\textbf{PM}} The $\sim$1.2 yr component is actually a doublet at 1.19 $\pm$ 0.00(4) and 1.20 $\pm$ 0.00(4) yr, known as the Chandler wobble (\cite{Chandler1891a, Chandler1891b}). The three components, Markowitz, Chandler and annual, amount respectively to 7.5, 40.4 and 19.8\% of the signal variance, for a total of 67.7\% \citep{Lopes2021}.\\

\begin{figure}[h!]
	\begin{subfigure}[b]{\columnwidth}
		\centerline{\includegraphics[width=\columnwidth]{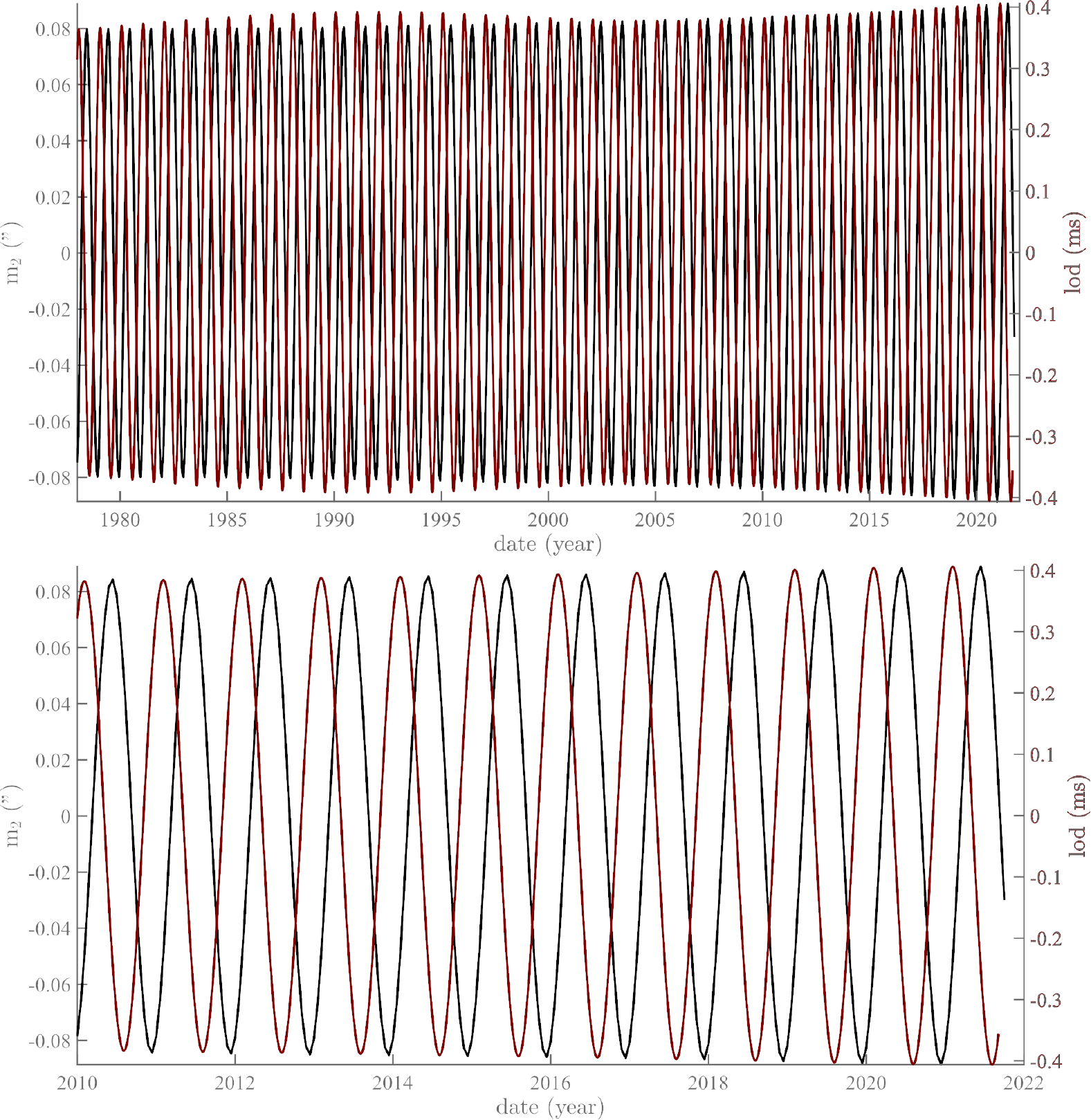}} 	
         \caption{Superimposition of the \textbf{SSA} annual components of the $m_2$ and \textbf{lod} time series (top: 1978-2022; bottom enlargement: 2010-2022)}
		\label{Fig:05a}
	\end{subfigure}
	\begin{subfigure}[b]{\columnwidth}
		\centerline{\includegraphics[width=\columnwidth]{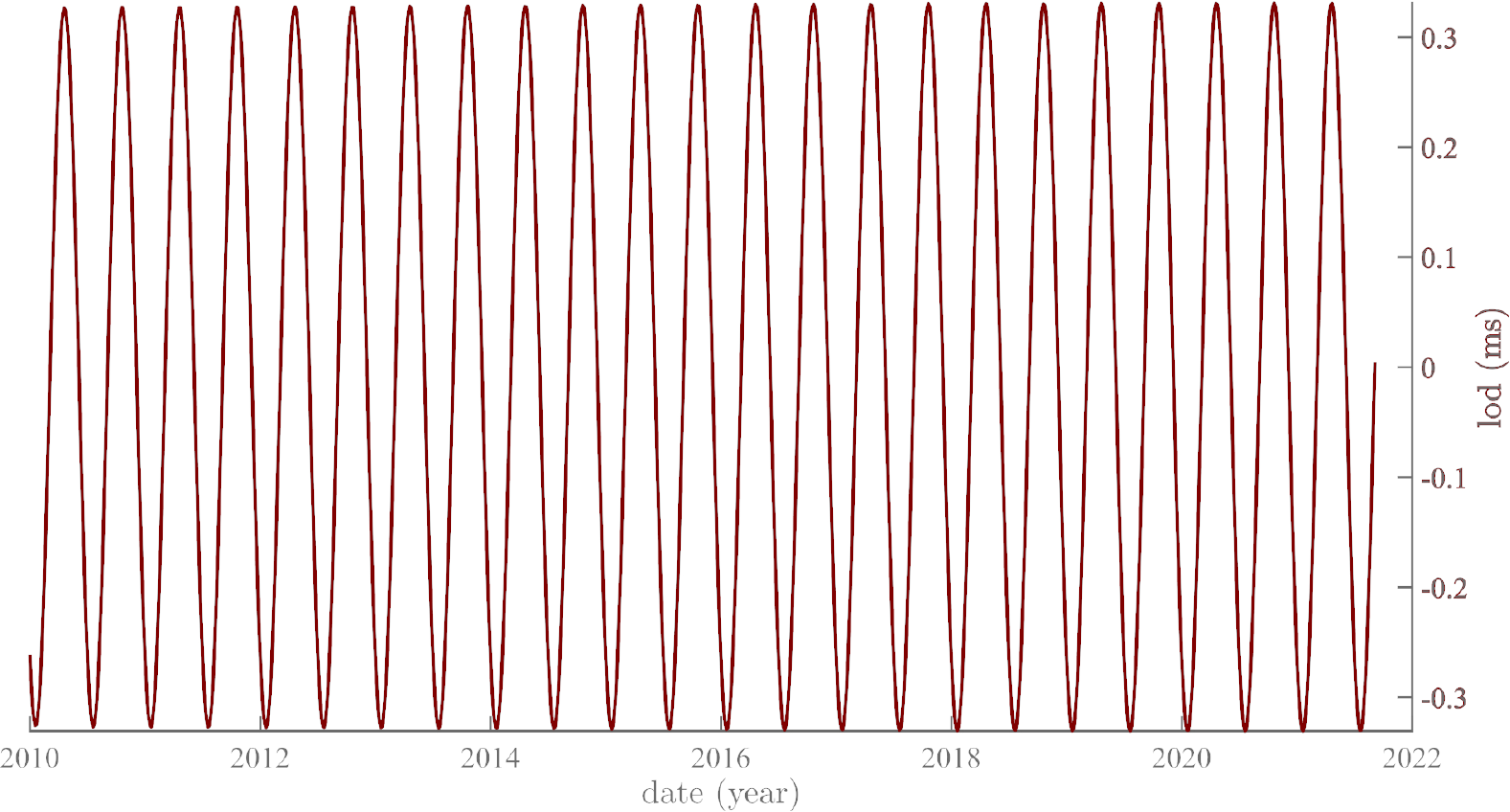}} 	
          \caption{\textbf{SSA} semi-annual component of the \textbf{lod} time series (2010-2022).}
		\label{Fig:05b}
	\end{subfigure}
	\caption{Annual and semi-annual components extracted from \textbf{lod} and \textbf{PM}}
	\label{Fig:05}
\end{figure}

Figures \ref{Fig:05a} and \ref{Fig:05b} compare the annual and semi-annual \textbf{SSA} components of \textbf{PM} and \textbf{lod}. The annual components are slightly modulated ($<$10\%) and are close to pure sinusoids. As a result, they almost perfectly correlate, with a phase difference of 41 $\pm$ 3 days between m2 and \textbf{lod} (\textit{i.e.} quadrature; Figure {\color{blue}{5}}). \\
\begin{table}[h!]
\centering
    \begin{tabular}{c c c c} 
         \hline
            & D$_{SE}$ & \textbf{\textbf{NHSI}} & \textbf{SHSI}\\ [0.5ex] 
         \hline
         \textbf{lod} & 152.5 $\pm$ 4.1 & 33.0 $\pm$ 1.1 & 153.7 $\pm$ 7.1 \\ 
         $m_{1}$ & 65.6 $\pm$ 2.1 & 153.8 $\pm$ 4.2 & 11.5 $\pm$ 2.4 \\
     \hline
     \end{tabular}
     \caption{Phase differences of the annual lines of the pairs given as column and line headings (in days, that is in this case very close to 1 day = 1 degree).}
      \label{Tab:01}
\end{table}

\begin{figure}[h!]
        \centerline{\includegraphics[width=\columnwidth]{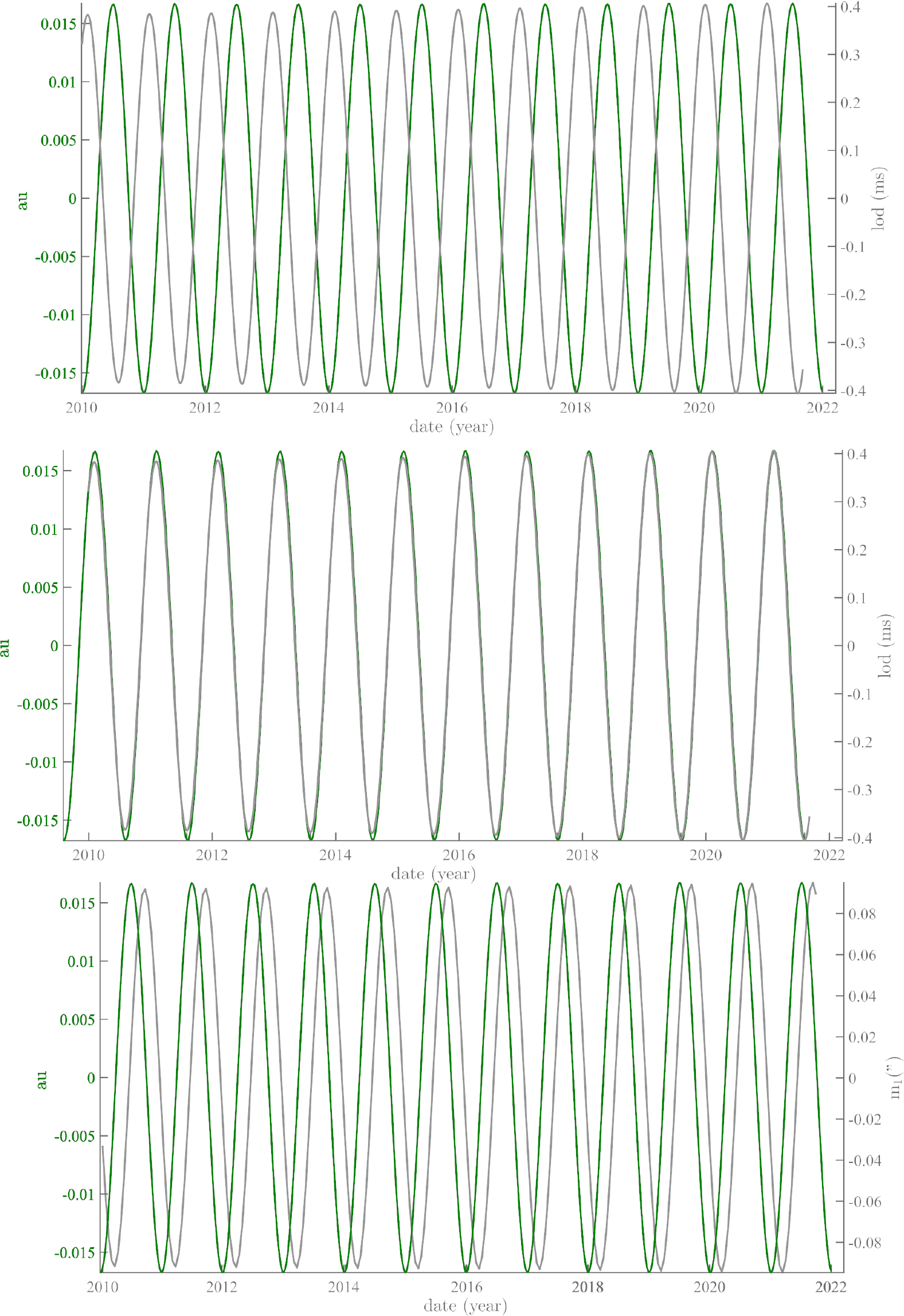}} 	
        \caption{Variations in the Earth to Sun distance D$_{SE}$ (in au; green) compared to the first (annual) \textbf{SSA} component of $m_1$ (bottom) and \textbf{lod} (top). In the middle curve, \textbf{lod} is offset by 152.5 days}
        \label{Fig:06}
\end{figure}

   The prominence of an “astronomical” annual component in these time series of course makes one relate it to the revolution period of Earth around the Sun. We compare the annual variations of the Sun-Earth distance D$_{SE}$ (in au) with m1 (Figure \ref{Fig:06}, bottom) and \textbf{lod} (Figure \ref{Fig:06}, top). The phase lags are respectively 65.6 $\pm$ 2.1 days for D$_{SE}$  and $m_1$ and 152.5 $\pm$ 4.1 days for D$_{SE}$ and \textbf{lod}.     

\section{Comparison of SSA annual and semi-annual components of polar motion and length of day vs variations in sea-ice extent}
\begin{figure}[h!]
        \centerline{\includegraphics[width=\columnwidth]{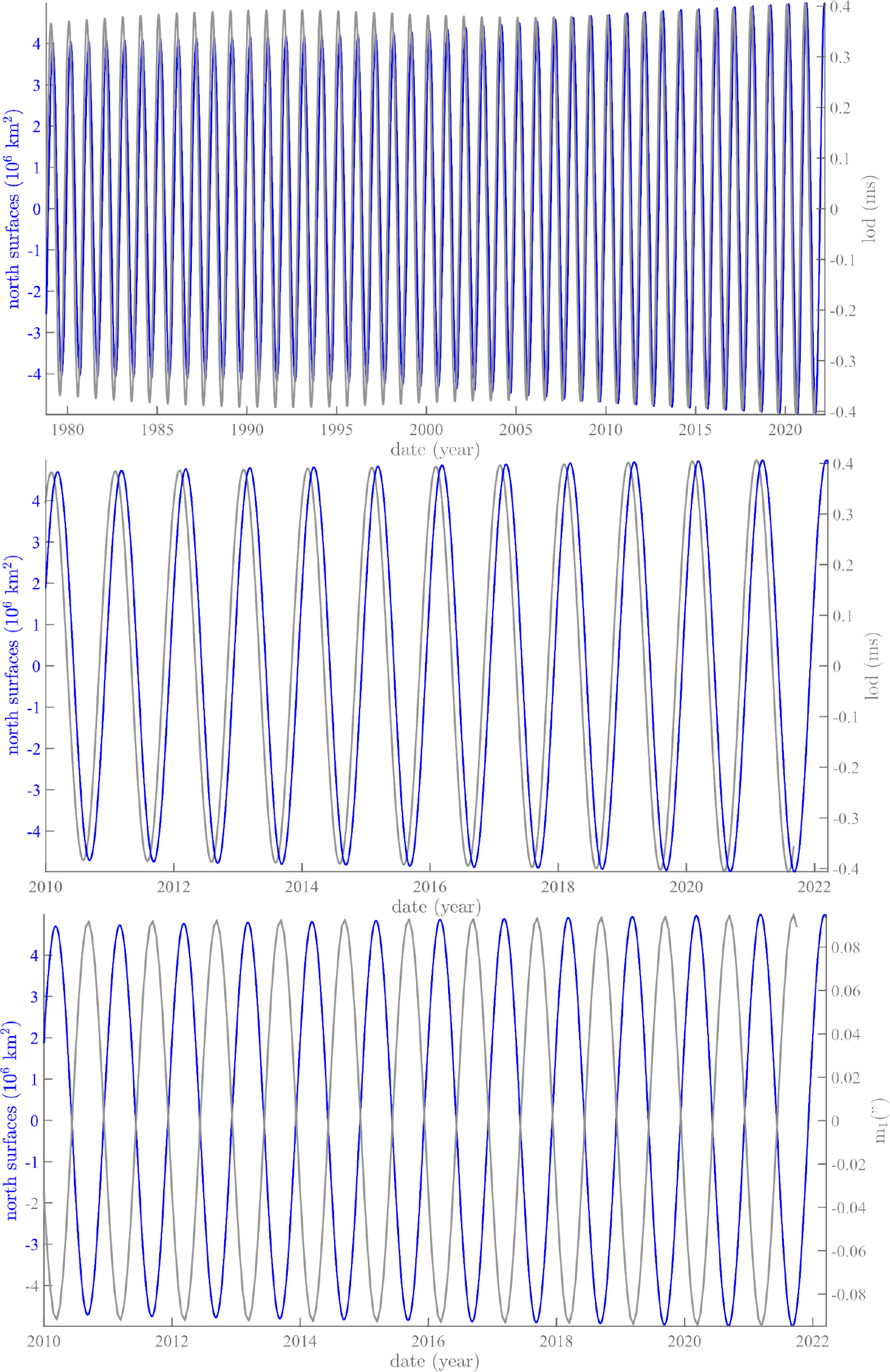}} 	
        \caption{Comparison of \textbf{SSA} annual components of pole motion $m_1$ (bottom; in arc seconds) and \textbf{lod} (top and middle enlargement; in ms) with that of \textbf{\textbf{NHSI}} (Arctic sea-ice; in 10$^6$ km$^2$). Top = 1978-2022; middle and bottom = 2010-2022.}
        \label{Fig:07}
\end{figure}

\begin{figure}[h]
        \centerline{\includegraphics[width=\columnwidth]{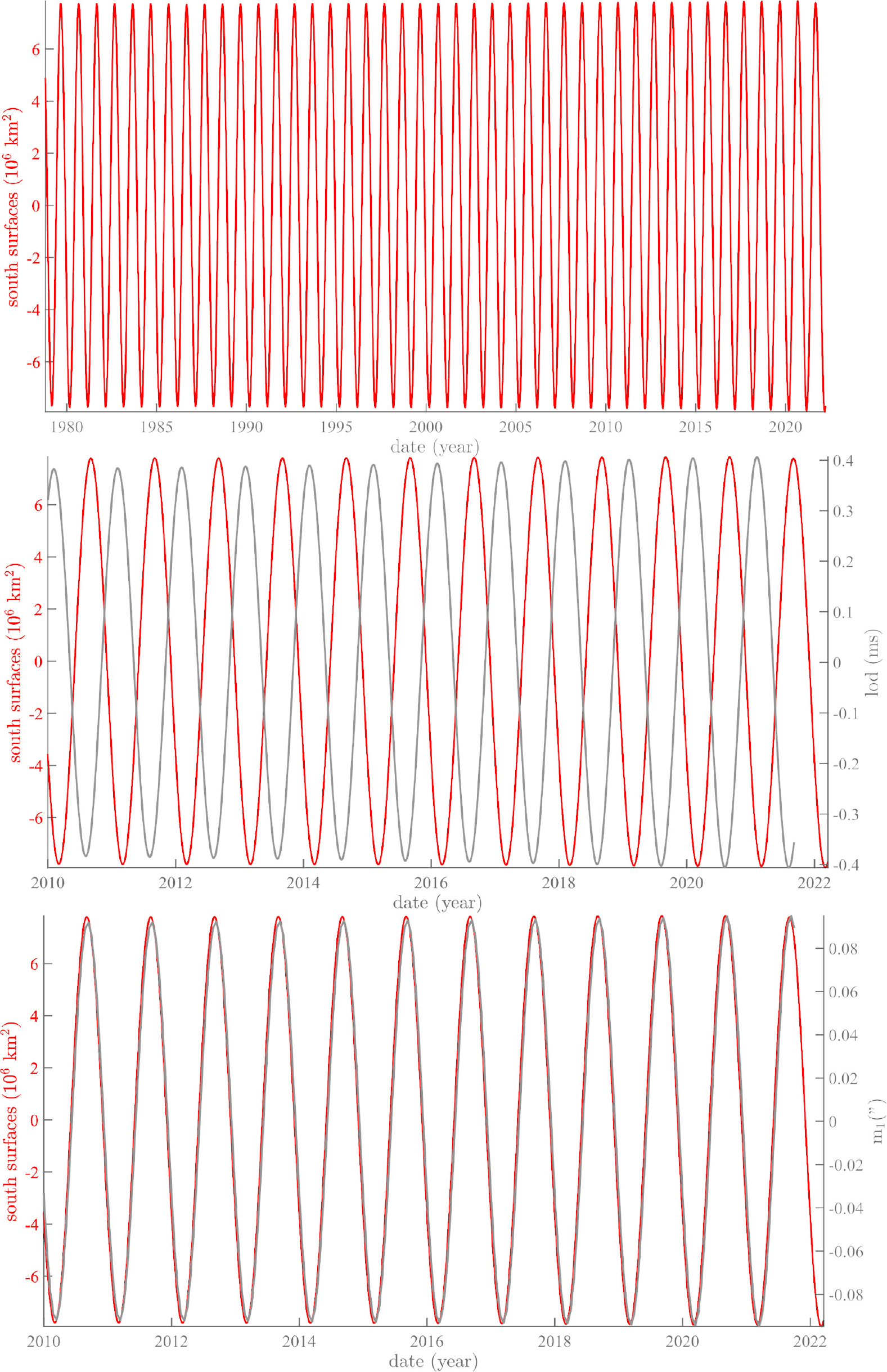}} 	
        \caption{ Comparison of \textbf{SSA} annual components of pole motion $m_1$ (bottom; in arc seconds) and \textbf{lod} (top and middle enlargement; in ms) with that of \textbf{SHSI} (Antarctic sea-ice; in 10$^6$ km$^2$). Top = 1978-2022; middle and bottom = 2010-2022.}
        \label{Fig:08}
\end{figure}

We now compare the annual components of sea-ice extent \textbf{NHSI} (Figure \ref{Fig:07}) and \textbf{SHSI} (Figure \ref{Fig:08}) with those of m1 and \textbf{lod}, in particular their phases (\ref{Tab:01}). In the case of the Northern hemisphere (Arctic), variations in sea-ice extent are almost in phase with \textbf{lod} and in phase opposition with $m_1$. The phase difference is 153.8 $\pm$ 4.2 days for m$_1$ and 33.0 $\pm$ 1.1 days for \textbf{lod} (see Figure \ref{Fig:07} middle enlargement).  \textbf{NHSI} is more strongly modulated (by 20\% over 40 years, corresponding mainly to the trend) than \textbf{lod}. In the case of the Southern hemisphere (Antarctic), variations in sea-ice extent are almost in phase with m1 and in phase opposition with \textbf{lod}. \\

\begin{figure}[h]
        \centerline{\includegraphics[width=\columnwidth]{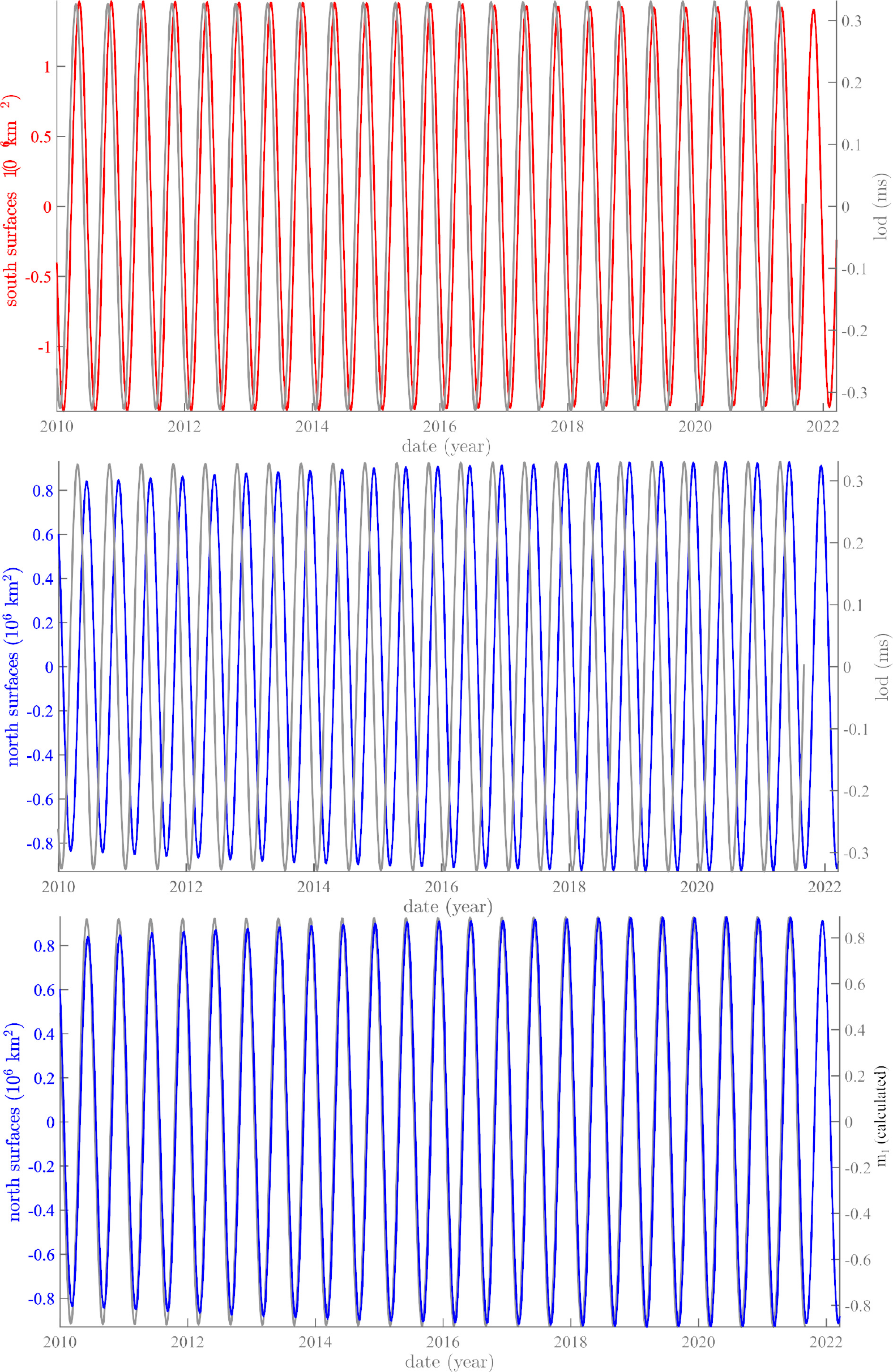}} 	
        \caption{Comparison of \textbf{SSA} semi-annual components of pole motion $m_1$ (bottom; in arc seconds) and \textbf{lod} (top and middle; in ms) with those of \textbf{\textbf{NHSI}} (middle and bottom; Arctic sea-ice; in 10$^6$ km$^2$) and of \textbf{SHSI} (top; Antarctic sea-ice; in 10$^6$ km$^2$) for 2010-2022. $m_1$ in the lower frame is not observed but is computed  by integrating \textbf{lod} (see text).}
       \label{Fig:09}
\end{figure}

The semi-annual component of \textbf{lod} is compared with that of \textbf{\textbf{NHSI}} in Figure \ref{Fig:07} (middle) and with that of \textbf{SHSI} in Figure \ref{Fig:09} (top). The respective phase differences are 51.8 $\pm$ 1.9 days (N) and 19.2 $\pm$ 1.5 days (S). Recall that the semi-annual component is present in \textbf{lod} but not in $m_1$. Thanks to the Euler-Liouville system (\cite{Laplace1799}), we can integrate the \textbf{lod} to obtain the theoretical polar motion that should accompany it. This is done in Figure \ref{Fig:09} (bottom). The agreement in phase is excellent. \\
    
 As shown in the previous section, these annual and semi-annual components of \textbf{NHSI} and \textbf{SHSI} are much larger (up to 20 times) than the trends observed since 1978. The mechanisms generally suggested as a forcing factor of the trends (global warming, winds,  see \cite{Zwally2002, Zhang2007,Turner2015,Parkinson2019}) must be dwarfed by a first order geophysical or astronomical forcing. Of course, the annual oscillating components of \textbf{NHSI} and \textbf{SHSI} are in opposite phases in both hemispheres (Figure \ref{Fig:02a} and \ref{Fig:02b}). \textbf{SHSI} is not significantly modulated whereas \textbf{NHSI} is. This is reminiscent of the classical behavior of a forced oscillator. Polar motion indeed possesses a forced annual component that is generally attributed to climate and ocean/atmosphere sources (\textit{e.g.} \cite{Jeffreys1916, Lambeck2005} chapter 7). \\

The lack of modulation of the largest (\textbf{SHSI}) forced component suggests an alternate mechanism. If we follow Laplace’s paradigm, the torques exerted by the Moon, Sun and planets play the leading role (if not all) as the source of forcing (modulation) of many (if not most) geophysical phenomena (generalized Earth tides). These forces (and torques) lead to changes in the inclination of the Earth’s rotation axis, transferring stresses to the Earth’s solid and fluid envelopes, setting Earth masses and resulting in thermal dissipation. This description is essentially that of \cite{Milankovic1920} on paleo-climates and ice ages: more than variations in eccentricity, it is variations in inclination of the rotation axis that lead to the large annual component of melting and re-freezing of sea-ice.\\

We have seen in Figure \ref{Fig:06} the excellent correlations of \textbf{lod} and $m_1$ with the Sun-Earth distance D$_{SE}$ (respectively in phase opposition and in phase. It is noteworthy that the phase differences of the annual components of \textbf{\textbf{NHSI}} with $m_1$, \textbf{SHSI} with \textbf{lod} and D$_{SE}$ with \textbf{lod} are all equal to 153 days, which happens to be half of the Euler period of 306 days (Table \ref{Tab:01}). \\
    
Let us follow Lambeck’s (1985, chapter 3, system 3.2.9) formulation of the Liouville-Euler equations (see also \cite{Lopes2021}):
\begin{equation}
	\begin{aligned}
		\dfrac{i}{\sigma_r} (\dfrac{d \textbf{m}}{dt}) + \textbf{m}&= \textbf{f} \\
         \dfrac{d m_3}{dt} &= f_3 \qquad (\textbf{lod})
	\end{aligned}
	\label{eq:01}
\end{equation}

$i=\sqrt{-1}$; $\textbf{m} = m_1 + i*m_2$ (polar coordinate). This is a linear first-order differential system. The annual oscillation of the forcing function $\textbf{f} = f_1 + i*f_2$ that follows sea-ice extent variations should behave as $m_1$ or $m_2$, without phase changes. This is indeed what is seen in Figures \ref{Fig:07},\ref{Fig:08},\ref{Fig:09} where phases remain constant. The fact that one recovers the phase differences of polar motion and \textbf{lod} in the leading annual components of sea-ice strengthens previous findings (\textit{e.g.} \cite{Courtillot2021, Lopes2021, Lopes2022a}). \textbf{lod} and $m_1$, $m_2$ are strongly linked.

\section{Discussion and Conclusion}
The differences between the frequency ranges of length of day (\textbf{lod}) and polar motion (\textbf{PM}) can be understood in the frame of \cite{Laplace1799} 's formulation of the Liouville-Euler system, which consists of an equation for polar axis inclination $\theta$ and one for the derivative of declination, $\dfrac{d\psi}{dt}$, both depending on $\theta$ only. The second equation that involves a derivative operator amounts to a high-pass filter\footnote{\scriptsize The derivative operator is the reason why \textbf{lod} has a richer content in higher frequency components than $m_2$.}, whereas the first behaves as an integrator, \textit{i.e.} passes lower frequencies. \\

Let us explore further consequences of Laplace’s formulation. We have seen that there is no semi-annual component in polar motion (\textit{e.g.} \cite{Lopes2021,Lopes2022a}. According to the theory of forced oscillators (\textit{e.g.} \cite{Landau1989}, the forcing frequency, here the annual oscillation, is at the boundary between the two frequency domains. Figure \ref{Fig:07} (center) shows that the semi-annual components of \textbf{NHSI} and \textbf{lod} are almost in quadrature (51.8 $\pm$ 1.9 days, close to 90/2 = 45 days for quadrature). Since there is a derivative operator between \textbf{lod} and \textbf{PM} and since from Figures \ref{Fig:05}, \ref{Fig:06} and \ref{Fig:07} the forced annual oscillations may be assumed to follow variations in the Earth to Sun distance D$_{SE}$, in \textbf{lod}, in \textbf{PM }and finally in \textbf{NHSI }and \textbf{SHSI} with remarkably constant phase differences, then any semi-annual oscillation of \textbf{lod} should theoretically also be present in \textbf{PM}. This would perturb polar motion and thus must be eliminated (dissipated), which is why we find it in \textbf{NHSI} and \textbf{SHSI}, and in quadrature with \textbf{lod} (Figure \ref{Fig:07}, bottom). \\
    
Integrating the semi-annual component of \textbf{lod} should give access to the same semi-annual component of $m_1$ that does not exist but should be in phase with the semi-annual component of \textbf{NHSI}, given the behavior of the ice series annual component. The result is shown in Figure \ref{Fig:07} (bottom) under the label “calculated $m_1$”. The phases and modulations of the two curves are the same (yielding in addition a prediction of the semi-annual component of \textbf{NHSI}). This is why all forced oscillations and their harmonics are found in variations in sea-ice, with very small amplitude modulations and remarkably stable phase differences between both hemispheres and with \textbf{lod} and \textbf{PM}. \\

The generally accepted origin of the annual oscillation in \textbf{PM} is the reorganization of fluid masses at the Earth surface. We may quote somewhat extensively the introduction of {\textcolor{blue}{Lambeck’s (2005) chapter 7}}:\\

 “\textit{The principal seasonal oscillation in the wobble is the annual term which has generally been attributed to a geographical redistribution of mass associated with meteorological causes. Jeffreys, in 1916, first attempted a detailed quantitative evaluation of this excitation function by considering the contributions from atmospheric and oceanic motion, of precipitation, of vegetation and of polar ice. Jeffreys concluded that these factors explain the observed annual polar motion, a conclusion still valid today, although the quantitative comparisons between the observed and computed annual components of the pole path are still not satisfactory. These discrepancies may be a consequence of (i) inadequate data for evaluating the known excitations functions, (ii) the neglect of additional excitation functions, (iii) systematic errors in the astronomical data, or (iv) year-to-year variability in the annual excitation functions}”. \\
 
\begin{figure}[h!]
	\begin{subfigure}[b]{\columnwidth}
		\centerline{\includegraphics[width=\columnwidth]{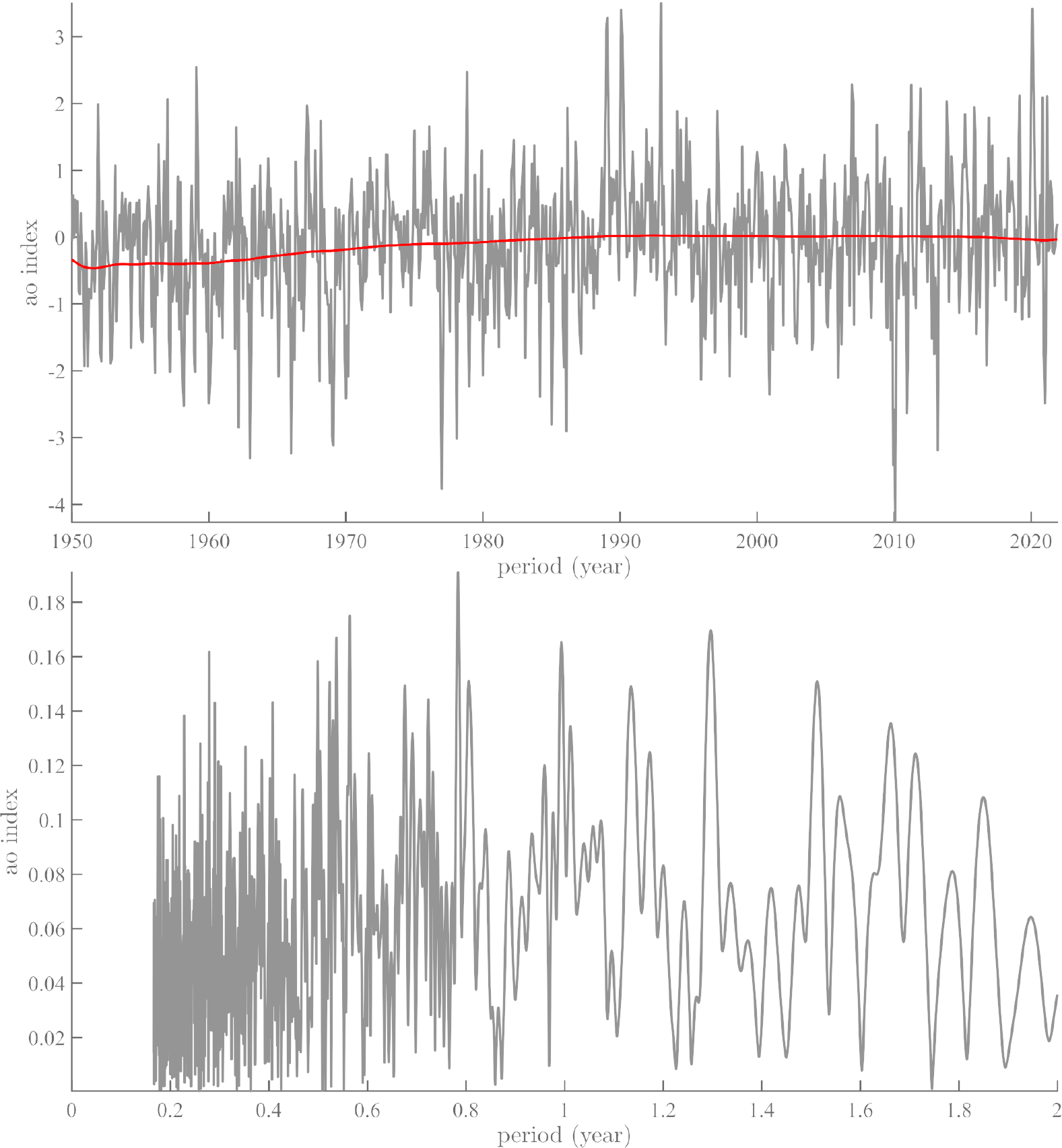}} 	
         \caption{AO Index, its mean (red) and spectrum (lower curve).}
		\label{Fig:10a}
	\end{subfigure}
	\begin{subfigure}[b]{\columnwidth}
		\centerline{\includegraphics[width=\columnwidth]{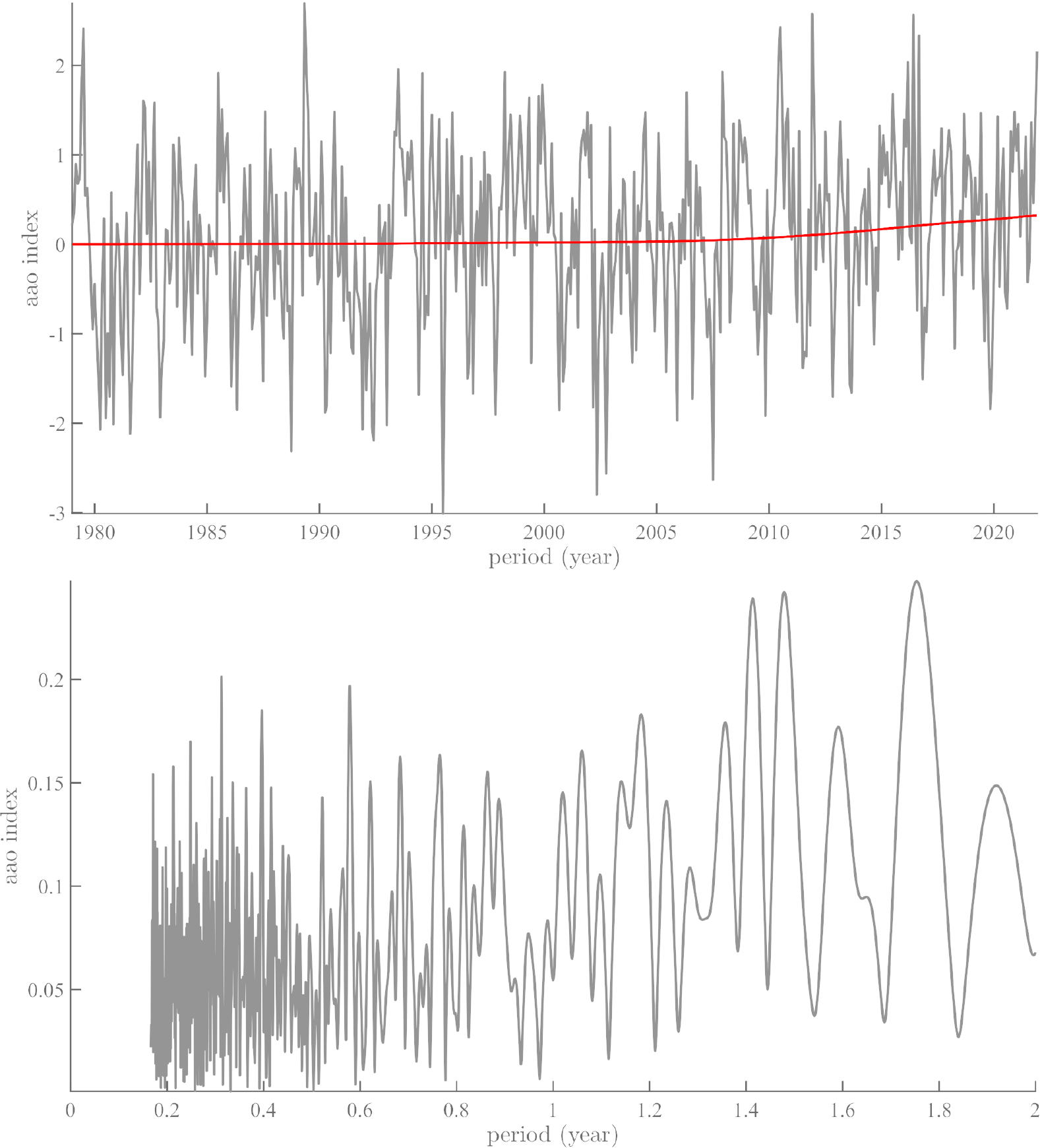}} 	
          \caption{AAO Index, its mean (red) and spectrum (lower curve).}
		\label{Fig:10b}
	\end{subfigure}
	\caption{AO and AAO Indices}
	\label{Fig:10}
\end{figure}

In other words, \cite{Lambeck2005} explains that the annual oscillation of \textbf{PM} is in part caused by variations in sea-ice and in part by variations of the atmospheric circulation. Regardless of whether one follows \cite{Laplace1799} or \cite{Jeffreys1916}, the observations do show remarkably constant phase differences. Yet, there is no reason for the periodic components of all climate indices to have the same relative or absolute phases (\textit{e.g.} \cite{LeMouel2019b}) (\textit{e.g.} Figures \ref{Fig:10a} and \ref{Fig:10b}).\\

    It is common to find in the literature suggestions that the evolution of sea-ice extent should be captured by climate indices, in the present case first and foremost the Arctic oscillation \textbf{AO} (\textit{cf.} Figure \ref{Fig:10a}, \cite{Thompson1998}) and the Antarctic Oscillation \textbf{AAO} (\textit{cf.} Figure \ref{Fig:10b}, \cite{Thompson1998}). The \textbf{AO} is characteristic of exchanges in atmospheric mass between the Arctic Ocean and the zonal ring centered on 45°N. \cite{Rigor2002} have for instance shown how changes in surface winds associated with the trends and fluctuations of \textbf{AO} influence sea-ice motion in the center of the Arctic, and how these changes in ice motion in turn influence the thickness and concentration of sea-ice, and the distribution of surface temperature above the Arctic \citep{Stroeve2007,Serreze2019, Cohen2020}. The \textbf{AAO} monitors the periodic reinforcement and weakening of the circumpolar vortex; it is also called the austral annular mode \textbf{AAM}. It is a stationary mode, unlike the Antarctic circumpolar waves \citep{White1996} that explain a large part of Antarctic climate variability on pluri-annual scales. The two \textbf{AO} and \textbf{AAO} indices are widely used in order to explain the evolution of sea ice since 1978. Yet, their spectra (Figures \ref{Fig:10a} and \ref{Fig:10b}) are devoid of any significant annual and semi-annual periodicities, when the \textbf{NHSI} and \textbf{SHSI} series display annual and semi-annual periodicities that amount to more than 98\% of the total signal variance. Nor are annual and semi-annual periodicities found in 10 \textbf{MJO} indices and several other index series, such as \textbf{PDO} and \textbf{ENSO} \citep{LeMouel2019b}. On the other hand, polar motion and length of day are global phenomena that could explain the regularity of phase and amplitude modulations of the sea-ice oscillating components at 1 and 0.5 year. \\
    
    Thus, the formulation by Laplace and the modern analysis of periodic and semi-periodic components of sea-ice variations are not that far from one another, although the direction of causality may have to be changed. If we follow Laplace’s view, variations in sea-ice extent are a consequence of polar motion as can be read from the Liouville-Euler equations. On the other hand, in the modern interpretation, they are consequences of wind speed variations, that should be reflected by indices \textbf{AAO} and \textbf{AO}. \\
    
    The results obtained in this paper on phases and amplitudes of \textbf{SSA} components of \textbf{NHSI} and \textbf{SHSI} allow one to prefer Laplace’s formulation. This formulation allows one to understand the presence of the harmonic series (1, 1/2 , 1/3, 1/4, 1/5, 1/6, 1/7, 1/8 yr) in Figure \ref{Fig:01b}. In a schematic way, variations in sea-ice extent are the infinitesimal (incremental) expression of oceanic motions, a topic that requires use of the theory of fluid mechanics of turbulent flow. There is still no solution of the problem in the spherical case \citep{Schrauf1986, Mamun1995, Nakabayashi1995, Hollerbach2006, Malhoul2016, Garcia2019, Mannix2021}. But there is one in the case of a flow between two coaxial cylinders \citep{Taylor1923, Landau1959,Chandrasekhar1961, Frisch1995}. The perturbation in flow velocity $v_1$ can be written as:
\begin{equation}
    v_1(r,\varphi,z) = f(r)*e^{i(n\varphi + kz - \omega t)}
    \label{eq:02}
\end{equation}
    
where integer $n$ is the order of symmetry ($n$=0 for axial symmetry) and $k$ is the wavelength of the instability. The acceptable values of the annual oscillation are obtained by solving equation \ref{eq:02} under the appropriate boundary conditions ($v_1$  = 0 for $r$ = $R_1$ and $r$ = $R_2$).  The Reynolds number $Re$ can be taken as  $\dfrac{\Omega_1R_1^2}{\nu}$ (or $\dfrac{\Omega_2R_2^2}{\nu}$ since $\dfrac{R_1}{R_2}$ and $\dfrac{\Omega_1}{\Omega_2}$ are given). For fixed $n$ and $k$ the values of $\omega$ form a discrete suite $\omega_n (k)$. By analogy with Earth, the solid surface and free surface of the atmosphere are such that $v_1 = 0$, and the only solutions for abs($\omega$) are 1 day and 1 year. We should therefore find harmonics of these two values. A detailed study of the solution of equation 02 in the case of co-axial cylinders is given by \cite{Chandrasekhar1961}. The $v_1$  flows are stationary toric vortices also known as Taylor vortices, regularly placed along the cylinder generatrices. In the case of two cylinders rotating in the same sense, one finds two vortices rotating in opposite directions, $2\pi/k_{cr}$, $k_{cr}$ being the critical wavelength above which $v_1$ is not 0. So, we now know how to describe the figure of the turbulent flow between two co-axial cylinders.\\

 Given the above, it is no surprise to find the series of harmonics $\omega_n (k)$ of the annual forcing. Also, the term $n\varphi$ in equation \ref{eq:02} forces the spatial symmetries of the flow. This is already known regarding the variations of the sea-ice dipole in Antarctica (\textit{cf.} \cite{Holland2017, Bertler2018, LeMouel2021b} and the variations of the trends of global atmospheric pressure (\textit{cf.} \citep{Lopes2022b}). These space-time variations seem to be better explained by Taylor-Couette type flow forced by variations in pole motion (rotation), basically as envisioned by \cite{Laplace1799}.

 \section*{Appendix A}
{\textcolor{blue}{Laplace (1799, vol. 5, cap. 1, page 347)}} in french:\\

 “\textit{\textcolor{darkgray}{Nous avons fait voir (n°8), que le moyen mouvement de rotation de la Terre est uniforme, dans la supposition que cette planète est entièrement solide, et l’on vient de voir que la fluidité de la mer et de l’atmosphère ne doit point altérer ce résultat. Les mouvements que la chaleur du Soleil excite dans l’atmosphère, et d’où naissent les vents alizés semblent devoir diminuer la rotation de la Terre: ces vents soufflent entre les tropiques, d’occident en orient, et leur action continuelle sur la mer, sur les continents et les montagnes qu’ils rencontrent, paraît devoir affaiblir insensiblement ce mouvement de rotation. Mais le principe de conservation des aires, nous montre que l’effet total de l’atmosphère sur ce mouvement doit être insensible; car la chaleur solaire dilatant également l’air dans tous les sens, elle ne doit point altérer la somme des aires décrites par les rayons vecteurs de chaque molécule de la Terre et de l’atmosphère, et multipliées respectivement par leur molécules correspondantes; ce qui exige que le mouvement de rotation ne soit point diminué. Nous sommes donc assurés qu’en même temps que les vents analysés diminuent ce mouvement, les autres mouvements de l’atmosphère qui ont lieu au-delà des tropiques, l’accélèrent de la même quantité. On peut appliquer le même raisonnement aux tremblements de Terre, et en général, à tous ce qui peut agiter la Terre dans son intérieur et à sa surface. Le déplacement de ces parties peut seul altérer ce mouvement; si, par exemple un corps placé au pole, était transporté à l’équateur ; la somme des aires devant toujours rester la même, le mouvement de la Terre en serait un peu diminué; mais pour que cela fut sensible, il faudrait supposer de grands changement dans la constitution de la Terre}}”. \\
 
 \section*{Appendix B}
\begin{figure}
\centerline{\includegraphics[width=\columnwidth]{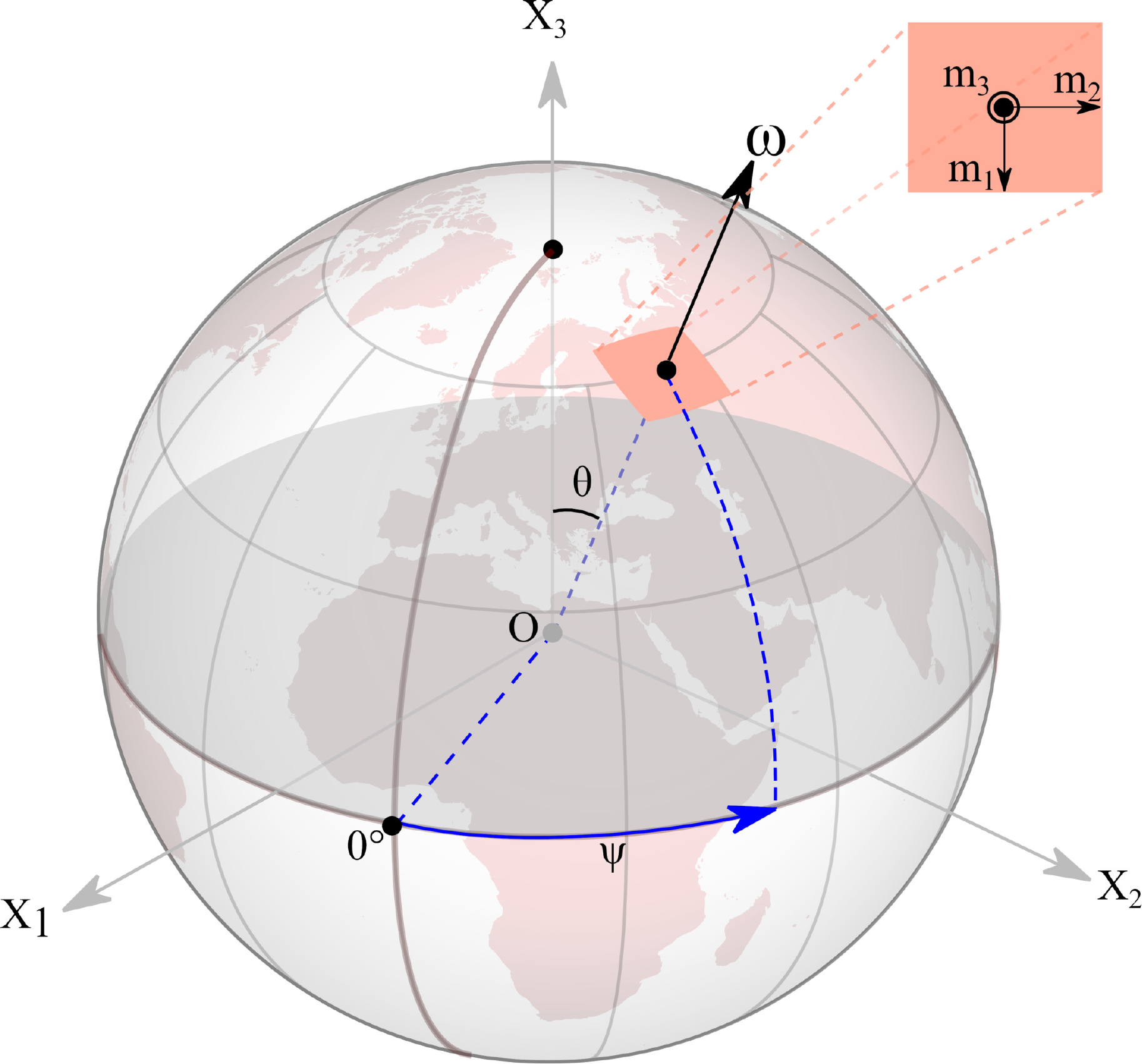}} 	
\caption*{Terrestrial reference frame. $m_1$ and $m_2$ are the coordinates of the rotation pole. $\psi$  and $\theta$  are the declination and inclination introduced by \citet{Laplace1799}. }
\label{Fig:B}
\end{figure}

Laplace gives the full equations with the Sun and Moon included: \\

\begin{subequations}
\begin{equation}
\theta = h + \dfrac{3m}{4n}.(\dfrac{2C-A-B}{C}) 
\left\{
 	\begin{array}{lll}
		    &\dfrac{1}{2}.\sin(\theta).[\cos(2\nu) + \dfrac{\lambda m}{m'}\cos(2\nu')] \\
        	& -(1+\lambda).m.\cos(\theta).\Sigma \dfrac{c}{f}.\cos(ft+\varsigma) \\
        	& + \dfrac{\lambda c'}{f'}.\cos(\theta).\cos(f' t + \varsigma')
 	\end{array}
\right.
	\label{eq:Ba}
	 \nonumber
\end{equation}

\begin{equation}
 \begin{array}{ll}
\dfrac{d \psi}{dt} &=  \dfrac{3m}{4n}.(\dfrac{2C-A-B}{C}) * \\
&\left\{ 
 \begin{array}{lll}
        & (1+\lambda).m.\cos(\theta) - \dfrac{\cos(\theta)}{2}.\dfrac{d}{dt}[\sin(2\nu) + \dfrac{\lambda m}{m'}.\sin(2\nu')]\\
        &  (1+\lambda).m.\dfrac{\cos^2(\theta)-\sin^2(\theta)}{\sin(\theta)}.\Sigma c.\cos(ft+\varsigma)\\
        &  \lambda .m\dfrac{\cos^2(\theta)-\sin^2(\theta)}{\sin(\theta)}. c'.\cos(f't+\varsigma')
    \end{array}
\right. 
 \end{array}
     \nonumber
	\label{eq:Bb}
\end{equation}
\end{subequations}

Note that $\theta$ and $\dfrac{d \psi}{dt}$  depend on $\theta$ but not $\psi$ .\\

The inclination $\theta$ of the rotation axis has the current value h in the first equation. $\dfrac{d \psi}{dt}$ is linked to the Earth's rotation, therefore to the \textbf{lod}. On the right side of the two equations are the ephemerids and masses of the Moon and Sun that enter the classical theory of gravitation (see Appendix A in \cite{Lopes2021} for more details). Length of day and polar inclination are clearly connected by the first relationship. Thus, Laplace reduces the problem to a system of two equations for the inclination and time derivative of the declination of the Earth's rotation axis.  $\theta$ and $\dfrac{d \psi}{dt}$ (and the norm that can be considered as a known constant) give the direction of the polar rotation axis and its variations. The time difference (in ms) between the theoretical and measured Earth rotation is proportional $\dfrac{\psi}{v}$, $v$ being the rotation velocity (and the Earth's radius is a constant).  Either  $\psi$ alone, or $v$ alone, or both can vary. We assume the former, since the mean rotation rate apparently remains constant, as was already noted above, and the two equation imply studying the time derivative of declination of the rotation axis, thus studying a quantity that is linearly related to the derivative of \textbf{lod}.
   
\newpage

\bibliographystyle{aa}

\end{document}